\documentclass[11pt,a4paper]{article}
\pdfoutput=1
\usepackage{amssymb}
\usepackage[numbers,sort&compress]{natbib}
\usepackage{amsmath,amsfonts,graphicx,epsfig}
\usepackage{bm}
\usepackage{caption}
\usepackage{xcolor}
\usepackage{amsmath,latexsym}
\usepackage{braket}
\usepackage{jcappub}

\def\bea{\begin{eqnarray}}
\def\eea{\end{eqnarray}}
\def\be{\begin{equation}}
\def\ee{\end{equation}}
\def\ba{\begin{array}}
\def\ea{\end{array}}

\def\P{{\mathcal P}}

\def\C{{\mathcal C}}

\def\fnl{{f_{\rm NL}}}

\newcommand{\overbar}[1]{\mkern 1.5mu\overline{\mkern-1.5mu#1\mkern-1.5mu}\mkern 1.5mu}

\title{Primordial black hole formation with non-Gaussian curvature perturbations}

\author[a]{Vicente Atal,} \author[a]{Jaume Garriga}\author[b,c]{and Airam Marcos-Caballero}

\affiliation[a]{Departament de F\'isica Qu\`antica i Astrofi\'sica, i  Institut  de  Ci\`encies  del  Cosmos, Universitat de Barcelona,\\Mart\'i i Franqu\'es 1, 08028 Barcelona, Spain.}
\affiliation[b]{Instituto de F\'isica de Cantabria, CSIC-Universidad de Cantabria\\Avda. de los Castros s/n, 39005 Santander, Spain.}
\affiliation[b]{Department of Theoretical Physics, University of the Basque Country UPV/EHU \\Avda. de los Castros s/n, 39005 Santander, Spain.}
\emailAdd{vicente.atal@icc.ub.edu, jaume.garriga@ub.edu, marcos@ifca.unican.es}

\abstract{ 
In the context of transient constant-roll inflation near a local maximum, we derive the non-perturbative field redefinition that relates a Gaussian random field with the true non-Gaussian curvature perturbation. Our analysis shows the emergence of a new critical amplitude $\zeta_*$, corresponding to perturbations that prevent the inflaton from overshooting the local maximum, thus becoming trapped in the false minimum of the potential. For potentials with a mild curvature at the local maximum (and thus small non-Gaussianity), we recover the known perturbative field redefinition.  We apply these results to the formation of primordial black holes, and discuss the cases for which $\zeta_*$ is smaller or of the same order than the critical value for collapse of spherically symmetric overdensities. In the latter case, we present a simple potential for which the power spectrum needs an amplitude 10 times smaller that in the Gaussian case for producing a sizeable amount of primordial black holes. 
}

\begin{document}

\maketitle

\section{Introduction} 

It is certainly an appealing possibility that black holes could have been formed primordially in an early period of evolution of our Universe \cite{Hawking:1971ei}. These primordial black holes (PBHs) could account for the dark matter in the Universe \cite{Carr:1974nx} and be the seeds of the supermassive black holes present in the center of most galaxies (for a recent review on the subject, see e.g. \cite{Sasaki:2018dmp}). If present, they could not only give an explanation for these intriguing observations, but would shed  light on the mechanism creating the primordial inhomogeneities. For example, if inflation is the theory accounting for those initial perturbations, then  the creation of PBH would require a far from canonical realization of inflation, possibly pointing towards its microphysical origin.

In order to confront the hypothesis of their existence with observations, we need to understand their formation process and statistical properties. If PBHs come from large fluctuations of the primordial density perturbations, and if those obey Gaussian statistics, then their description is well known \cite{Bardeen:1985tr}. Gaussianity  of the primordial perturbations might seem to be supported by the fact that during canonical and slow-roll single-field inflation departures from Gaussianities are known to be small \cite{Maldacena:2002vr}. However, most models of PBH creation during inflation require a large peak in the power spectrum, and this amplification necessarily demands for departures from slow-roll \cite{Motohashi:2017kbs}\footnote{A notable exception, where no amplification of the power spectrum is needed, are models where PBH are formed from the collapse of domain walls created during inflation \cite{Garriga:2015fdk,Deng:2016vzb}.}. Most calculations so far on the impact on the non-Gaussianity  in the computation of abundances has been performed assuming slow-roll \cite{Bullock:1996at,Pattison:2017mbe}, or considering arbitrary templates for the shape of the non-Gaussianity \cite{PinaAvelino:2005rm,Young:2013oia,Young:2014ana,Young:2014oea,Young:2015cyn}. When calculations have been done in the transient constant-roll (CR) background, usually a perturbative approach has been followed, focusing on the impact of the three-point function of the curvature perturbation\footnote{The amplitude of non-Gaussianities in transient periods of CR is different from their amplitude in a pure CR \cite{Martin:2012pe,Namjoo:2012aa,Bravo:2017wyw}. The difference between both scenarios is that in the former the curvature perturbations eventually freeze.}\cite{Saito:2008em, Cai:2017bxr,Atal:2018neu,Passaglia:2018ixg}. In this case it has been shown that the curvature mode follows the so-called local template of non-Gaussianities, which mean that the non-Gaussian field $\zeta$ can be written in terms of a Gaussian field $\zeta_g$ through the following simple transformation law
\be\label{eq:zeta_local_trans}
\zeta_g \rightarrow \zeta=\zeta_g + \frac{3}{5}\fnl \zeta_{\rm g}^2 \ .
\ee
In single-field models of inflation the parameter $\fnl$ can be calculated, and it is given by \cite{Cai:2017bxr,Atal:2018neu}
\begin{align}
\fnl&=\frac{5}{12}\left(-3+\sqrt{9-12\eta}\right) \\
&=\frac{5}{12}\left(-6+\epsilon_2^{cr}\right)
\end{align}
\noindent where $\eta\equiv V''/V<0$ is the curvature of the inflaton potential at the top of the local maximum, and $\epsilon_2^{cr}$ is the second slow roll parameter evaluated in the constant-roll period\footnote{We call constant-roll whenever $\epsilon_2\leq-6$, thus incorporating ultra slow-roll ($\epsilon_2=-6$).}. In this scenario the cosmological background interpolates between the constant-roll background and its dual graceful exit ('slow-roll') background, described by $\epsilon_2^{ge} = -6+\epsilon_2^{cr}$ \cite{Wands:1998yp,Kinney:2005vj,Tzirakis:2007bf,Morse:2018kda,Atal:2018neu}. The amplitude of $\fnl$ can then be written as
\begin{align}
\fnl&=\frac{5}{12}\epsilon_2^{ge} \ .
\end{align}
Because after overshooting a local maximum the field evolves, at least during a small period of time, in a large curvature potential, \emph{{\`a} la} hill-top, $\epsilon_2^{ge}$ can naturally be $\mathcal{O}(1)$. This bump could possibly set the end of inflation (because the initial velocity is exponentially small after climbing the potential, it is still possible to get a reasonable number of e-foldings), or it could be matched to a potential with a smaller curvature, delaying the end of inflation.

The above relation shows that $\fnl$ can be $\mathcal{O}(1)$, and this turns out to be large enough to invalidate the Gaussian assumption for all single-field models of inflation claiming to produce PBHs. This has been shown in \cite{Atal:2018neu} by applying the criteria  proposed in \cite{Franciolini:2018vbk}. Now, the expression (\ref{eq:zeta_local_trans}) suggests that this is just the first order term of a general field redefinition, only valid when $\fnl \zeta \ll 1$. The large values of $\fnl$ and $\zeta$ that describe most models of PBH formation suggest that this truncation is not justified. A full non perturbative description of the non Gaussianities is then important. As we will show in the section \ref{sec:NP} of this paper, the complete field redefinition is
\be\label{eq:zetanptransf}
\zeta = -\frac{1}{\beta}\ln\left(1-\beta\zeta_g \right)
\ee
\noindent with $\beta=6/5\fnl$. For small values of $\beta$, the local  transformation law (\ref{eq:zeta_local_trans}) is recovered.\\

Aside from this intrinsic non-Gaussianity, resulting from the microphysics of inflation, large curvature perturbations are non-linearly related to the density fluctuations. This effect can be taken into account by noting that $\zeta$ is simply related in superhorizon scales to the so-called compaction function $\C(r)\equiv \delta M/R$  with $\delta M$ the mass excess over the FRW background, $R$ the areal radius and spherical symmetry has been assumed. At superhorizon scales this function is time independent and takes the following form\footnote{At the radius where this function is maximal, $r=r_{\rm m}$, this function is related to the average density perturbation $\delta_{\rm m}\equiv\delta(r=r_{\rm m})$  \cite{Harada:2015yda}.}
\be\label{eq:compac}
\C(r)=\frac{1}{3}\left(1-\left(1+r\partial_r\zeta\right)^2\right) \ .
\ee
If at early times this function has an initial amplitude above a certain threshold, then a black hole is formed \cite{Shibata:1999zs}. This means that it is possible to find the associated threshold for the curvature mode bypassing the need of calculating the density perturbation at all scales (for which a linearised relation with the curvature mode is usually assumed). This criteria for PBH formation, and its impact in the calculation of abundances, was followed in \cite{Yoo:2018esr}. Quantitatively, larger amplitudes of the power spectrum, by a factor $\sim 2$,  are needed for reaching a desired abundances of PBH in comparison with the linear theory \cite{Kawasaki:2019mbl,Young:2019yug,DeLuca:2019qsy}.\\

In this work we will show, on the one hand, that when $\fnl \sim 1$ the amplitude of the power spectrum must be smaller by a similar factor compared to the case with $\fnl=0$, in order to attain the same abundace of PBH. On the other hand, and because of the non-pertubative law (\ref{eq:zeta_local_trans}), we will show that for large values of $\fnl$ perturbations will rather jump into the false minimum of the potential than create spiky peaks. Black holes will still be created, but this time through the collapse of bubbles of false vacuum \cite{Garriga:2015fdk,Deng:2016vzb}.

\section{The non-perturbative field redefinition}\label{sec:NP}

\begin{figure}[t!]
\begin{center}
\includegraphics[scale=0.6]{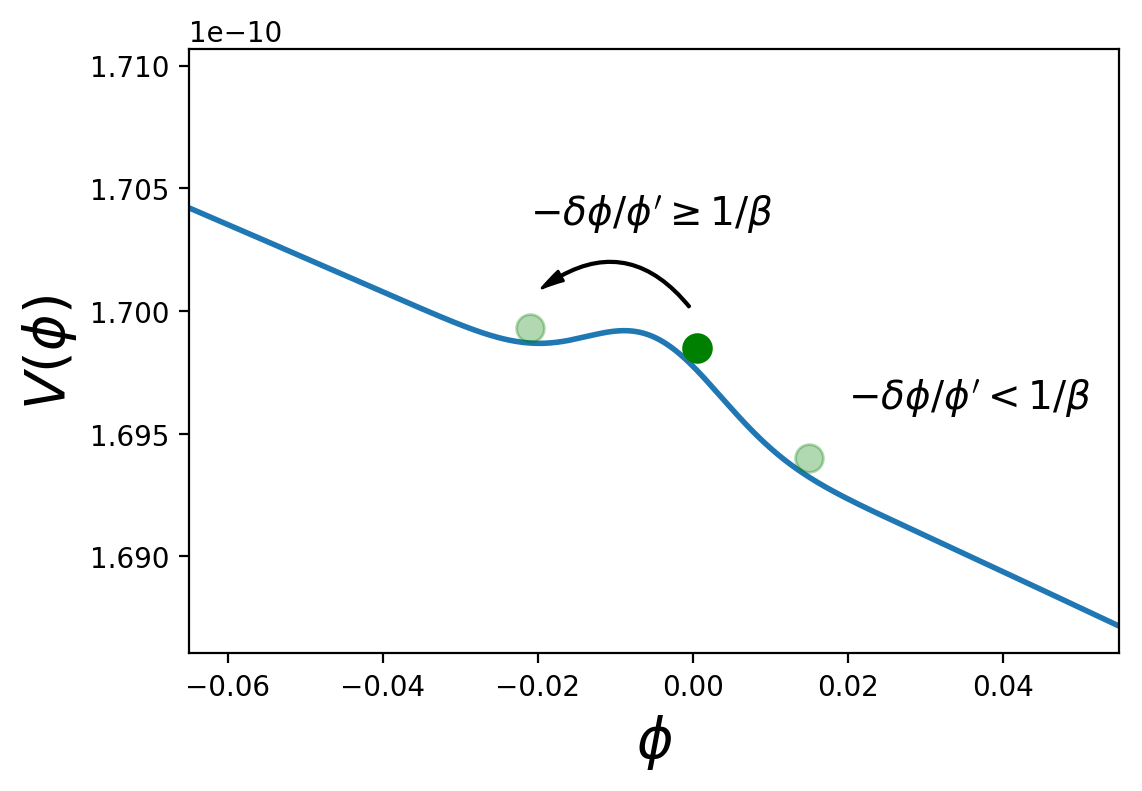}
\caption{The potential introduced in eq. (\ref{eq:pot_staro}) (with shifted values for $\phi$), where we describe the consequences of large fluctuations on top of the background trajectory. The background solution overshoots the local maximum. Fluctuations larger than $1/\beta$ at horizon crossing  make the field become trapped in the false minimum of the potential, while those smaller than $1/\beta$ do not.}
\label{fig:potbump}
\end{center}
\end{figure}

Single-field and canonical models of inflation can experience an enhancement of the power spectrum at small scales if the background trajectory passes through a phase of constant-roll. Many single-field models accomplishing this with different microscopic motivations have been put forward in the literature \cite{Yokoyama:1998pt,GB,Germani:2017bcs,Ballesteros:2017fsr,Ozsoy:2018flq,Cicoli:2018asa,Dalianis:2018frf}. However they all belong to a very simple class of models in which the background overshoots a local maximum. It is then enough to consider the following expansion of the potential around this maximum, i.e.
\begin{equation}
V \left(\phi\right)=\frac{\eta}{2}(\phi-\phi_c)^2 
\end{equation}
\noindent where we consider $\eta<0$. In order to find the non perturbative relation between the Gaussian and the non-Gaussian field we can make use of the $\delta N$ formalism\cite{Starobinsky:1986fxa}. Here we closely follow the analysis of ref. \cite{Cai:2017bxr}\footnote{For a pure USR, i.e., without a smooth transition to a graceful exit phase, see \cite{Namjoo:2012aa,Franciolini:2018vbk}.   }. The homogeneous solution in terms of the e-folding number $dN= H dt$ is
\begin{equation}\label{eq:back}
\phi(N)=\phi_c+c_1\,e^{\alpha N} + c_2 \,e^{\beta N} 
\end{equation}
\noindent with $\alpha=\frac{1}{2}\left(-3-\sqrt{9-12\eta}\right)$ and $\beta=\frac{1}{2}\left(-3+\sqrt{9-12\eta}\right)$. Note that $\alpha<0$ and $\beta\geq0$, so at late times the solution proportional to $c_2$ dominates. The constants $c_1$ and $c_2$ are functions of the initial velocity and position, given by
\begin{align}
c_1&=\frac{1}{\alpha-\beta}\left[\phi'_i+\beta\left(\phi_c-\phi_i\right)  \right] \ , \\
c_2&=\frac{1}{\beta-\alpha}\left[\phi'_i+\alpha\left(\phi_c-\phi_i\right)  \right] \ . \label{eq:c21}
 \end{align}
Without loss of generality, we can consider the inflaton field evolving from negative to positive values, so that $\phi_i'>0$ and $\phi_c-\phi_i>0$. The field overshoots the local maximum of the potential if at late times it goes towards $+\infty$, which means that $c_2$ must be positive. On the other hand, if $c_2=0$ the field stops exactly at the local maximum and if $c_2<0$ the field bounces back. We can now evaluate the number of e-foldings from an initial time $N=0$ until the the beginning of the slow-roll regime, which we call $N_{\rm SR}$. The slow-roll regime  is the regime in which the growing solution dominates, and we can always choose this to happen at a value of $\phi=0$. In this case we find
\be
c_2\simeq -\phi_c \,e^{-\beta N_{\rm SR}} \ .
\ee
\noindent We can now determine how a change in the initial conditions changes the number of e-foldings to reach $\phi=0$. We find
\be
\frac{\delta c_2}{c_2} =-1 + e^{-\beta \delta N|_{\phi=0}}  \ .
\ee
Note that there is no constraint on the amplitude of $\delta c_2$ with respect to $c_2$. On a flat slicing at $N=N_{\rm SR}={\rm const}$, $\delta\phi_{\rm SR}=\delta c_2 e^{\beta N_{\rm SR}}$. From (\ref{eq:back} we also have that $c_2=\phi'_{\rm SR} e^{-\beta N_{\rm SR}}/\beta$,  so
\be\label{eq:deltac2}
\frac{\delta c_2}{c_2}=\beta\frac{\delta \phi_{\rm SR} }{\phi'_{\rm SR}} \ .
\ee
This mean that the Gaussian variable $\zeta_g\equiv -\delta \phi_{\rm SR}/\phi'_{\rm SR}$ is related to the real non-Gaussian perturbation $\zeta\equiv\delta N$ as
\be
\zeta_g = \frac{1}{\beta}\left(1- e^{-\beta \zeta}  \right) \ .
\ee
By inverting this relation we find eq. (\ref{eq:zetanptransf}).
 As we have already anticipated, this identification has several interesting features. Curvature perturbations are not well determined when the Gaussian variable has amplitudes larger than $1/\beta$. Physically these are perturbations that prevent the system of overshooting the local maximum. Given a background $c_2$, any perturbation $\delta c_2$ such that $\delta c_2 \leq -c_2$ will make $c_2\leq 0$, preventing the field from overshooting the local maximum. From (\ref{eq:deltac2}) this corresponds to the case $\zeta_g \geq 1/\beta$. We depict these two regimes in figure \ref{fig:potbump}, where we show the fate of fluctuations larger than $1/\beta$, that make the field become trapped in the false minimum, and fluctuations smaller than $1/\beta$ for which the trajectory of the field towards to global minimum is not altered. Regions that stay in the false minimum will form bubbles of false 
vacuum, and the fate of those is eventually to create black holes. Depending on their initial size, they might be standard black holes coming from the ordinary collapse of the bubble at subhorizon scales, or become black holes with an inflating internal region, that is, with a baby universe in their interior \cite{Garriga:2015fdk,Deng:2016vzb}.

We end this section with two comments. First of all, the amplitude of $\beta$ can be written in terms of second slow-roll parameter $\epsilon_2$ at the graceful exit phase. It holds that
\begin{align}
\beta&=\frac{1}{2}\left(-3+\sqrt{9-12\eta}\right) \\
&=\frac{\epsilon_2^{ge}}{2} \ .
\end{align}
Finally let us note that peaks of $\zeta_g$ are transformed into peaks of $\zeta$. Therefore we can perfectly count the number of peaks in the Gaussian variable to count the peaks in the non Gaussian one. 

\section{The typical high peak profile}

The computation of the threshold for collapse as well as the abundances of peaks are usually calculated using the mean profile of a peak as a representative realisation of their full distribution. The mean profile of a peak with amplitude $\mu$, averaged over all the possible values for curvatures at the peak, is given by a integral of the form
\be\label{eq:mean}
\overbar{F(r)}=\int \zeta P(\zeta | \rm{peak})d\zeta
\ee
\noindent where $P(\zeta | \rm{peak})$ is the conditional probability of $\zeta$ given that the field has certain amplitude $\mu$ at the position of the peak $r=0$. If the probability distribution function is Gaussian and if the peak is large, in the sense that the amplitude $\mu$ is much larger than the variance $\sigma_0$, i.e, $\nu\equiv \mu /\sigma_0 \gg 1$, then the mean profile of the peak is simply given by
\be\label{eq:meangaus}
\overbar{F(r)}=\mu \psi(r)  \ , 
\ee
where $\psi(r)$ is the correlation function $\psi(r)\equiv \langle F(r)F(0)\rangle$, given by
\be
\psi(r)=\frac{1}{\sigma_0^2}\int \frac{dk}{k}\frac{\sin(kr)}{kr}P_k
\ee
and $P_k$ is the power spectrum of the fluctuations. The variance of the  shape is given by
\be\label{eq:deltagaus}
\frac{\left(\Delta F\right)^2}{\sigma_0^2}=1-\frac{\psi^2}{1-\gamma^2}-\frac{1}{\gamma^2\left(1-\gamma^2\right)}\left(2\gamma^2\psi+\frac{R_s^2\nabla^2\psi}{3}\right)\frac{R_s^2\nabla^2\psi}{3}-\frac{5}{\gamma^2}\left(\frac{\psi'}{r}-\frac{R_s^2\nabla^2\psi}{3} \right)^2-\frac{\psi'^2}{\gamma^2}  \ ,
\ee
where $\gamma\equiv\sigma_1^2/(\sigma_2\sigma_0)$, $R_s\equiv \sqrt{3}\sigma_1/\sigma_2$, with $\sigma_n$ the $n$-moments of the distribution \cite{Bardeen:1985tr}. The mean profile coming from the non-Gaussian distribution found in the previous section is not well defined, since the integral (\ref{eq:mean}) diverges when $\zeta\rightarrow 1/\beta$. This is not an impediment for analyzing how do most profiles look like. Most of the realisations will have a Gaussian shape in between $\overbar{F}\pm \Delta F$, and so we only need to see how these shapes are deformed by the presence of non-Gaussianity. The deformation of the Gaussian shapes will depend on the amplitude of the profile, but we only want to focus on profiles that could possibly collapse to form BH. Here the identification  between $\zeta(r)$ and the compaction function $C(r)$ is helpful since the initial values of $C(r)$ leading to gravitational collapse are very well constrained. It holds that \cite{Kopp:2010sh,Harada:2013epa}\footnote{The dependence of the threshold on the shape of the initial density contrast has been extensively studied (see e.g. \cite{Musco:2018rwt}), as well as its impact on the abundances of PBH \cite{Germani:2018jgr}. For an exploration of these dependences on the local shape of non-gaussianity, see \cite{iliaprep}.}
\be\label{eq:C_cr}
0.21< C_{\rm th}(r_{\rm m}) < 1/3 \ ,
\ee
where $r_{\rm m}$ is the radius at which  $C(r)$ is maximum. The bound above can then be easily translated into a bound on the amplitudes of $\zeta$ at the peak, once a profile for $\zeta$ is given. This allows one to make reasonable predictions for the abundance of PBH even in the case in which there is no known exact value for the thresholds to collapse\footnote{By reasonable we mean that it only translates into a small uncertainty in the amplitude of the primordial power spectrum in order to reach a certain abundance. Of course, a small variation of the threshold value keeping the amplitude of the power spectrum fixed results in exponentially large uncertainties.}, as was for example done  in \cite{Atal:2018neu}. On the other hand, a comparison of the possible values for collapse evaluated at the center of the profile leads to unbounded uncertainties.   

For the power spectrum of the curvature perturbations, that determines the correlation function $\psi$, we consider the following: from CMB scales up to scales where the power spectrum grows we take a constant with amplitude $A_0\sim 10^{-9}$. We model the growth of the power spectrum with a scale dependence $k^4$. As argued in \cite{Byrnes:2018txb}, this is the maximum possible growth of the power spectrum. Most models actually saturate this bound during large part of their growth. The decay of the power spectrum after the peak on the other hand can be related to the amplitude of the non-Gaussianity \cite{Atal:2018neu}. The power spectrum can then be modelled as 
\begin{equation}
  \P(k) = 
   \begin{cases}
    A_0  & \text{for } k < k_{0} \\
     \P_0 \left(\frac{k}{k_{\rm p}}\right)^{4}  & \text{for } k_0 \leq k \leq k_{\rm p} \\
   \P_0 \left(\frac{k}{k_{\rm p}}\right)^{-n}  & \text{for } k \ge k_{\rm p}\ ,
   \end{cases}\label{eq:model_pw}
\end{equation}
where $\P_0$ is the amplitude at the peak of the power spectrum. The spectral index at scales smaller than $k_{\rm p}$, $n$,  is given by 
\be\label{eq:spec}
n\simeq-3+\sqrt{9-12\eta_V}.
\ee
By comparing (\ref{eq:spec}) with the expression for $\beta$, we see that both are related through 
\be\label{eq:beta_vs_n}
\beta\simeq\frac{n}{2} \ .
\ee
The abundance of peaks diverges when the power spectra decays slower than $k^{-3}$, since they depend on the second momenta $\sigma_2$ \cite{Bardeen:1985tr}. Thus, in order to make finite predictions we need to use a window function, which can be done by considering the coarse grained power spectrum $\P(k,k_{\rm cut})$ given by  
\be
\P(k,r_{\rm f}) \rightarrow W(k,r_*)\P(k)\ .
\ee
The choice of the window function might lead to uncertainties \cite{Ando:2018qdb}, although small \cite{Young:2019osy}. For simplicity, and because we concentrate on the dependence of the non-Gaussianity on the abundances,  we choose the top-hat window function in momentum space. We typically need $A_0\ll \P_0$ in order to have a significant abundance of PBH, so we can neglect the power at scales larger than $k_0$. That is, we finally consider a power spectrum given by
\begin{equation}
  \P(k,k_{\rm cut}) = 
   \begin{cases}
    0  & \text{for } k < k_{0} \\
     \P_0 \left(\frac{k}{k_{\rm p}}\right)^{4} \ , & \text{for } k_0 \leq k \leq k_{\rm p} \\
   0 \ , & \text{for } k > k_{\rm p}\ .
   \end{cases}\label{eq:model_pw}
\end{equation}
where we choose the coarse graining scale $k_{\rm cut}$ to be the scale of the peak of the power spectrum $k_{\rm p}$, since at this scale the abundace of PBH will be larger. The correlation function determining the shape of the peak is then simply given by
\be
\psi_g(r)\simeq\frac{4}{k_{\rm p}^4r^4}\left[-2+\left(2-k_{\rm p}^2r^2\right)\cos\left(k_{\rm p}r\right)+2k_{\rm p}r\sin\left(k_{\rm p}r\right) \right]
\ee
where we have further assumed that $k_0\ll k_{\rm p}$.

\begin{figure}[t!]
\begin{center}
\includegraphics[scale=0.5]{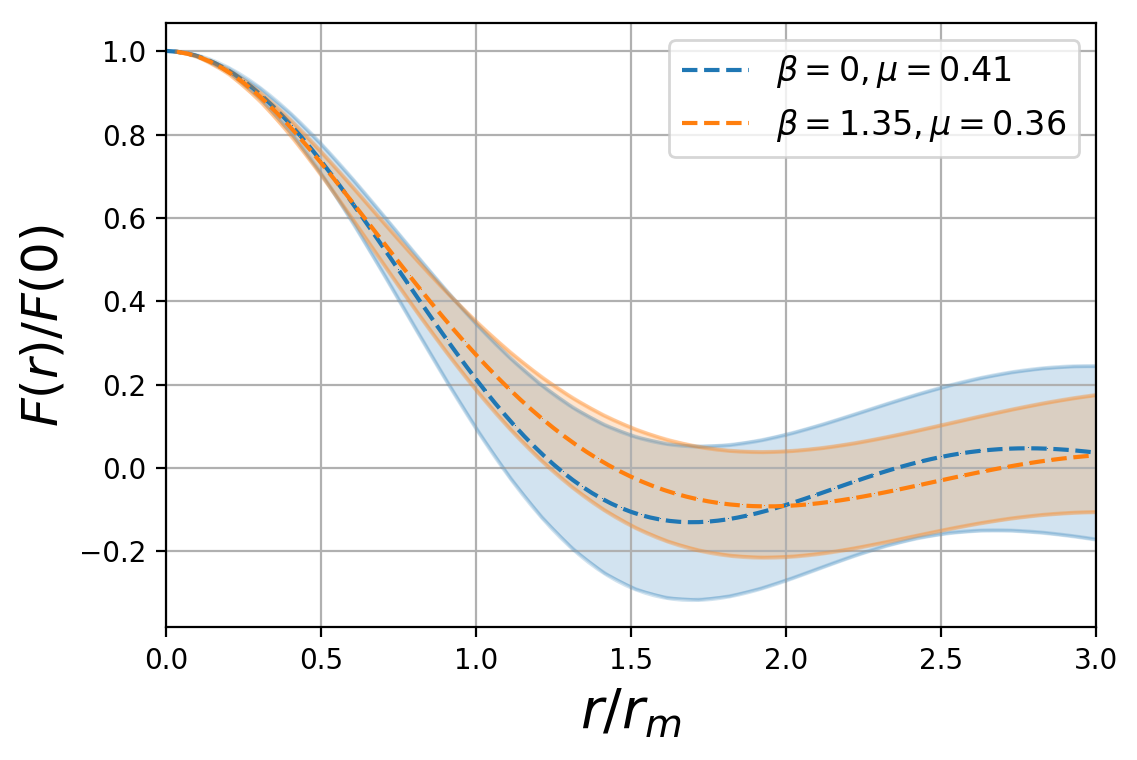}
\includegraphics[scale=0.5]{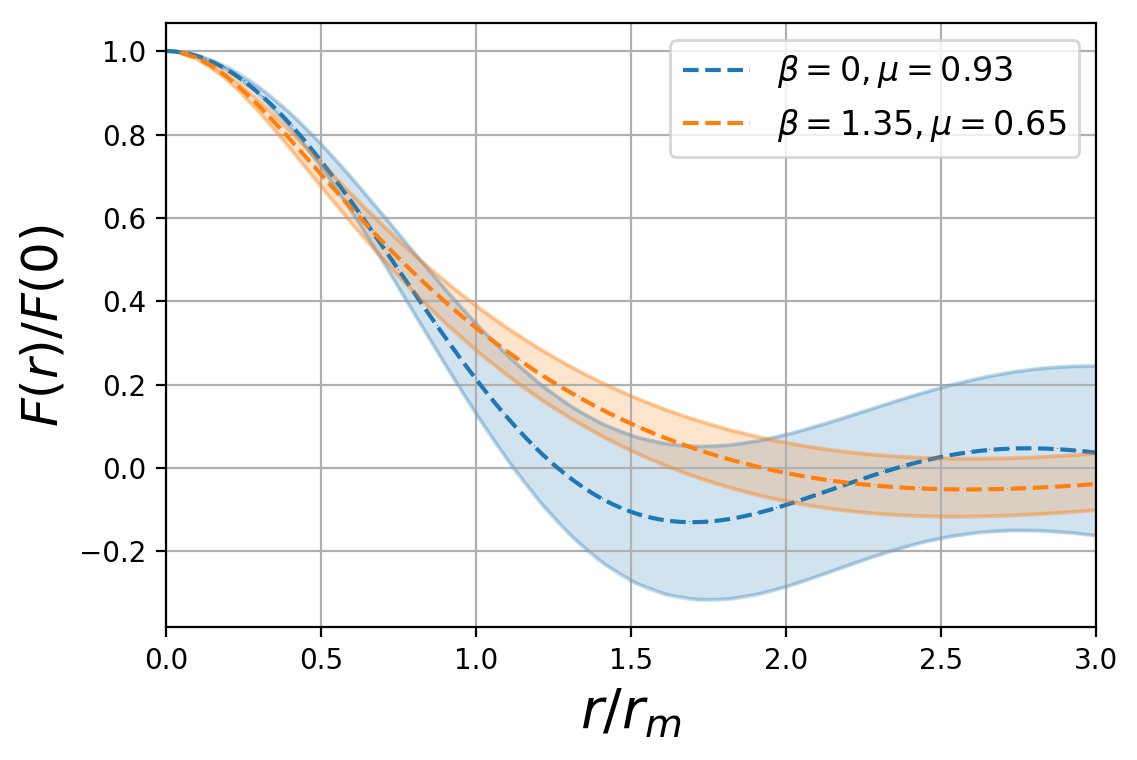}
\includegraphics[scale=0.5]{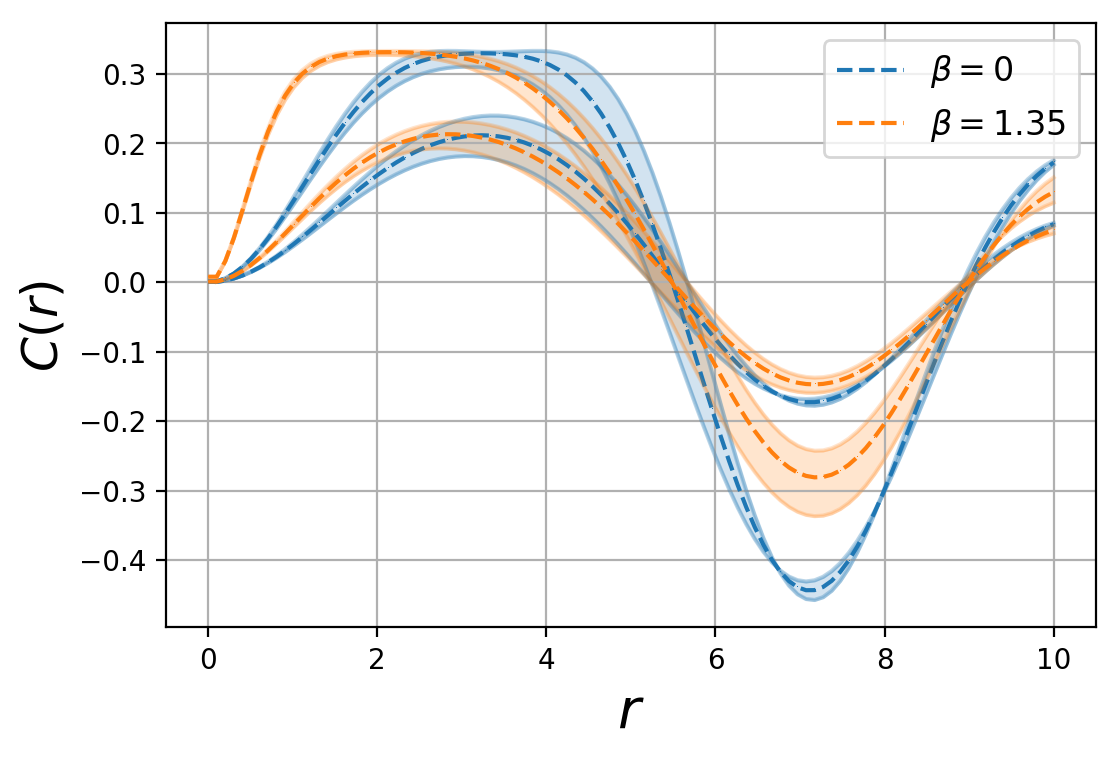}
\caption{\emph{top)} The mean and 1-$\sigma$ contours of the  Gaussian profiles and how they transform under non Gaussianity ($\beta=1.35$). For the non-Gaussian profiles these corresponds to the median and the 16\% and 84\% quartiles. We vary the amplitude $\mu$ so that profiles span all possible values for the compaction function that could in principle form PBHs, limits given by eq. (\ref{eq:C_cr}). We keep $\nu=5$ fixed. \emph{bottom)} Compaction functions in the critical range for the Gaussian and non Gaussian case (corresponding to the profiles shown above).}
\label{fig:profiles_ng}
\end{center}
\end{figure}

In figure \ref{fig:profiles_ng}  we show how the Gaussian set of profiles is transformed by the presence of non-Gaussianity. The non-Gaussian set of profiles is found by applying the transformation law (\ref{eq:zetanptransf}) to all Gaussian profiles ranging in the interval $\overbar{F_g}\pm \Delta F_g$, with $\overbar{F_g}$ and $\Delta F_g$ given by (\ref{eq:meangaus}) and (\ref{eq:deltagaus}) respectively. We take a fiducial value of $\beta=1.35$ (corresponding to $\fnl=1.25$), and vary the amplitude of the profile such that the compaction function lies in the critical range (\ref{eq:C_cr}). We also take\footnote{Larger values of $\nu$ will have smaller dispersion in their profiles and so in this sense we are consider the largest possible dispersion.} $\nu=5$. Interestingly, the dispersion of the profiles that could possibly collapse is systematically smaller for the non-Gaussian case. The reason is twofold. First, the radius $r_{\rm m}$ is smaller for the non-Gaussian case (and close to the center of the profile the dispersion is smaller). Secondly and most importantly, large value of the compaction function are attained for smaller values of $\mu$, and thus, if we keep $\nu\equiv \mu / \sigma$ fixed (since it is the most important parameter controlling the abundances), the dispersion $\sigma$ will be smaller. 
From here we deduce that while it is not possible to define a mean profile, it is certainly possible to define a typical profile, given in this case by the median profile of the distribution, which will be simply
\be
F_{\rm ng}=\frac{1}{\beta}\ln\left(1-\beta \overbar{F}_{\rm g}\right) \ .
\ee
In the next section we anayse the implications of such profile in the formation and abundances of PBHs.

\section{Implications for PBH }

The implications of the non-Gaussianity in the abundance of PBH will depend on whether the critical value for spherical collapse ($\zeta_c$) is smaller or of the same order than the critical value for staying in the false minimum of the potential $(\zeta_*$). In the following we will separately discuss both cases.

\subsection{Perturbative limit $\zeta_c < \zeta_*$}

We first consider the perturbative limit, for which the critical value for collapse ($\zeta_c$) is smaller than amplitude to end in the false minimum ($\zeta_*$). We take $\beta=1.25$, and the same critical value for the compaction function for both the Gaussian and non-Gaussian profiles, since the normalized profiles do not vary substantially with respect to each other for such values of $\beta$. We take the threshold to be $C_{\rm th}=0.255$ since the profiles are similar to those already studied in the literature, and for which the threshold has been determined \cite{Germani:2018jgr}. For the Gaussian case this amplitude of the compaction function is attained for $\mu=0.54$. In the non-Gaussian case, this threshold is attained for $\mu=0.44$. In order to keep the value of $\nu$ constant, the power spectrum needs an amplitude $\sim 1.5$ smaller in the non-Gaussian case. For this amplitude of $\beta$, this is an effect of a similar order than the one coming from the non-linear relation between $\zeta$ and $\delta_m$, but acts in the opposite direction. In the next section, we will see that this effect can be largely enhanced when the theory enters in the non-perturbative regime.

\subsection{Non-perturbative limit $\zeta_c \simeq \zeta_*$}\label{sec:PBHImp}

\begin{figure}[t!]
\begin{center}
\includegraphics[scale=0.5]{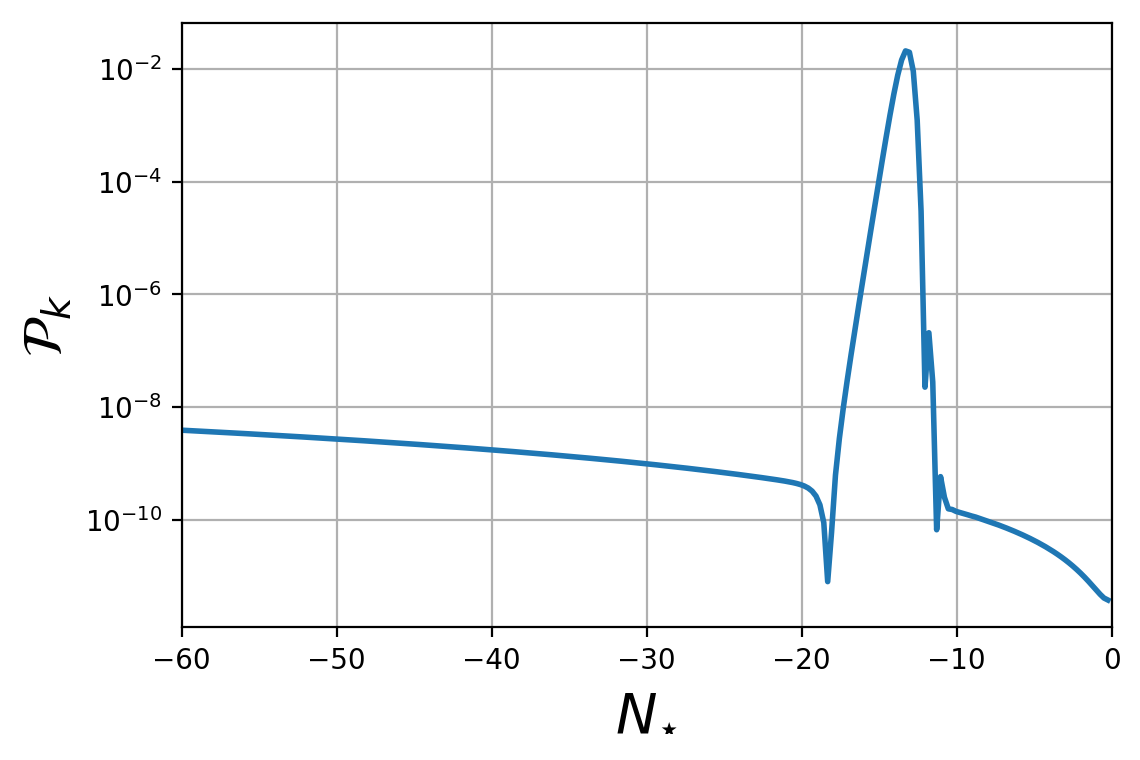}
\includegraphics[scale=0.5]{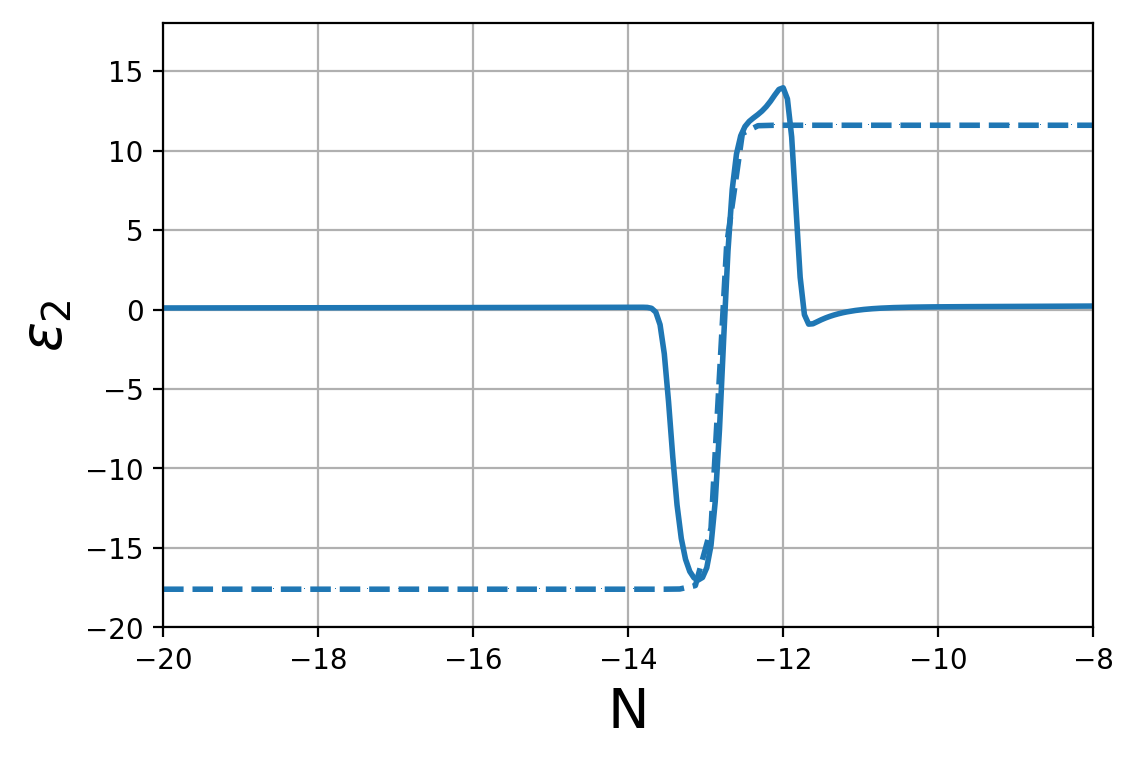}
\caption{\emph{left)} Power spectrum for the model given by the potential (\ref{eq:pot_staro}), together with the following parameters: $\{A=1.80325\times 10^{-3},\phi_0=3.7,\sigma^2=9\times 10^{-5}\}$. \emph{right)} Evolution of the second slow roll parameter $\epsilon_2$ as a function of time (measured in e-foldings $N$). The dashed line corresponds to $\epsilon_2$ coming from the solution (\ref{eq:back}) \cite{Atal:2018neu}. We see that it describes well the transition from the CR to the graceful exit phase.}
\label{fig:staro}
\end{center}
\end{figure}

\begin{figure}[t!]
\begin{center}
\includegraphics[scale=0.5]{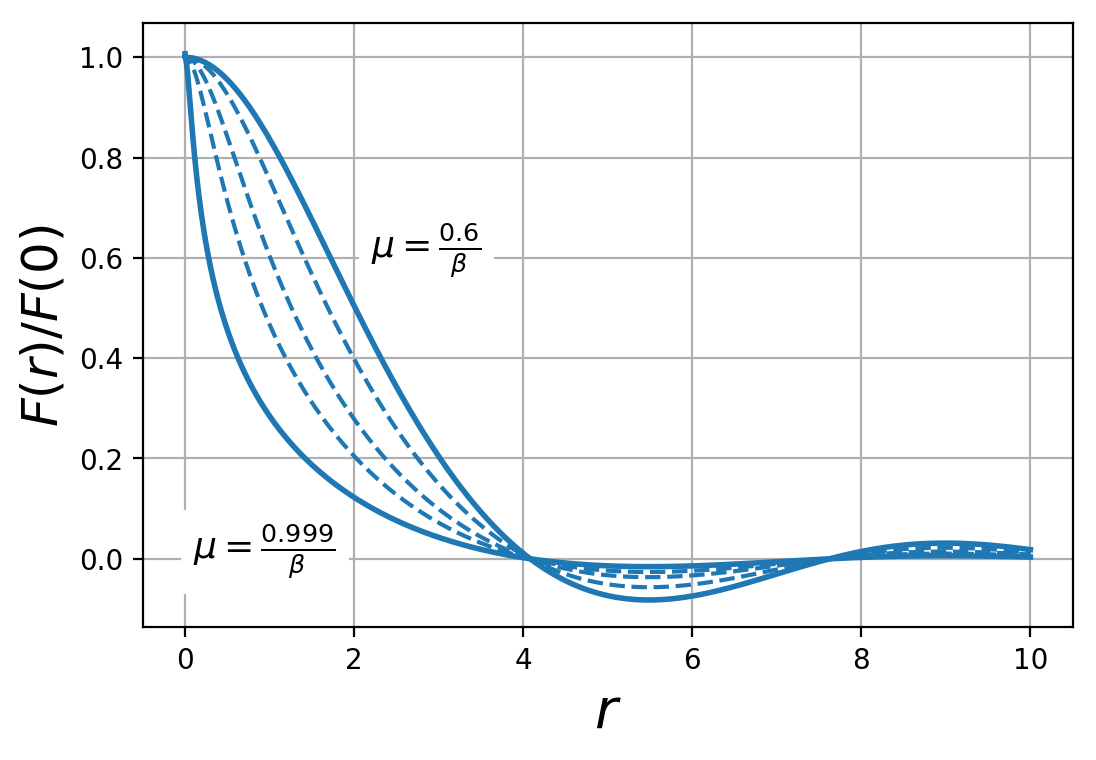}
\includegraphics[scale=0.5]{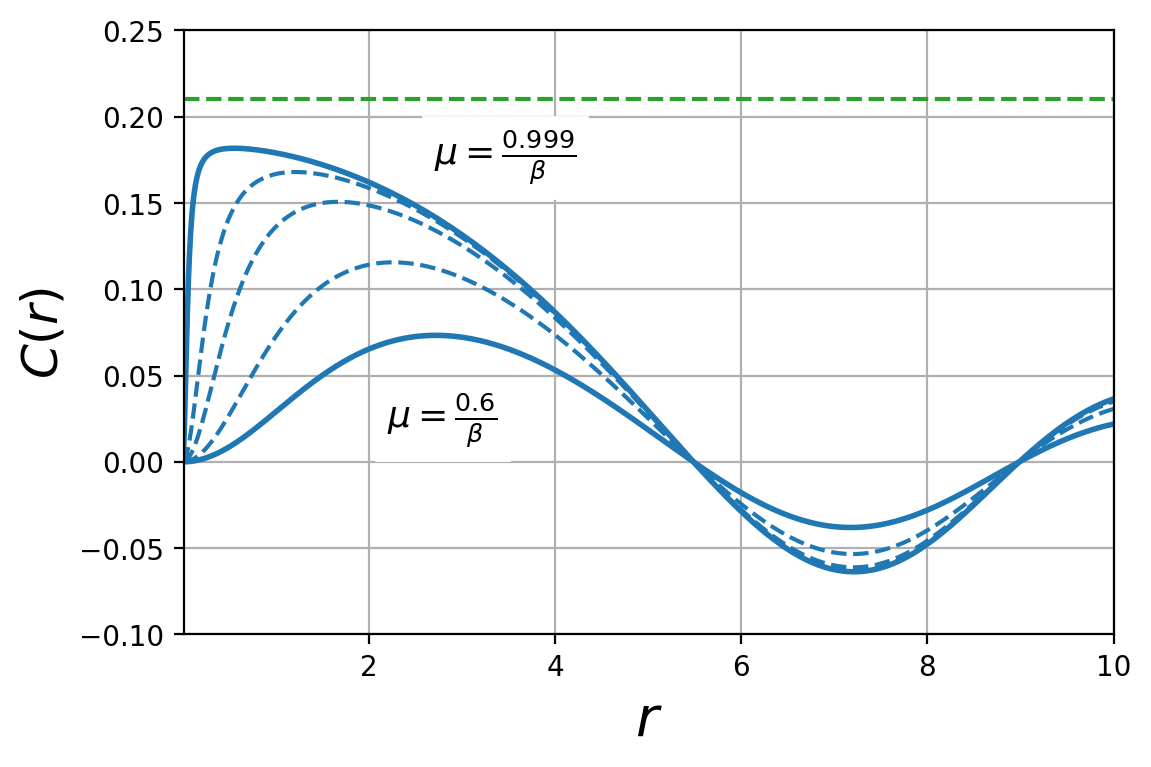}
\caption{The variation of the median profile and median compaction function in the non-Gaussian case, with $\beta=6$. While the amplitude is very close to the non-perturbative bound $1/\beta$ ($\mu=0.999/\beta$), the maximum of the compaction function is still lower than the lowest possible threshold for collapse, $C_{\rm th}=0.21$.}
\label{fig:prof_comp_npng}
\end{center}
\end{figure}

Now we consider the non perturbative limit for which $\zeta_c \simeq \zeta_*$. For this, larger values of $\beta$ should be attained, larger than in current models in the literature. We propose a simple model where this is realised. We consider the Starobinsky  potential \cite{Starobinsky:1980te} with a small bump given by a Gaussian function, i.e.
\be\label{eq:pot_staro}
V(\phi)=V_0^4\left(1-e^{-\sqrt{2/3}\phi}\right)^2\left[1+Ae^{-\left(\phi-\phi_0 \right)^2/\left(2\sigma^2\right)}  \right] \ .
\ee
In figure \ref{fig:staro} we show the power spectrum as well as the evolution of the second slow-roll parameter $\epsilon_2$, which determines the amplitude of $\beta$. 
The predicted value for $\epsilon_2$ at the graceful exit period is $\epsilon_2^{ge}=12$, and so\footnote{As can be seen in the figure, the actual value is slightly larger. This means that $f_{\rm NL}$ will correspondingly be slightly larger. Note that the fact that there is a large evolution of $\epsilon_2$ after the transition does not affect the amplitude of $\fnl$ at scales corresponding to the transition, where the peak of the power spectrum is, since perturbations freeze when $\epsilon_2=\epsilon_2^{ge}$ \cite{Atal:2018neu}.} $\beta=6$. In figure \ref{fig:prof_comp_npng} we show the profile as well as the compaction function as a function of the amplitude $\mu$. We see that for values of $\mu$ very close to the non perturbative critical value $1/\beta$, in this particular case $\mu=0.999/\beta$, the compaction function is still below the lowest of the standard critical threshold for the compaction function, namely $C_{\rm th}=0.21$. Note that if we choose amplitudes $\mu$ even closer to the threshold $1/\beta$, lets say $\mu=c/\beta$ with $c>0.999$, the compaction function will eventually rise above its threshold for collapse. However, the contribution of these fluctuations to the total abundance of PBH will correspond to the integration of the probability density function  in a very narrow range of the gaussian variable, that is from $c/\beta$ to $1/\beta$, where in this case $0.999<c<1$. On the other hand, the contributions from fluctuations that jump into the false minimum comes from fluctuations in the range $1/\beta$ to $\infty$, which will then dominate the production of PBH.

Finally let us note that because the inflaton field traverses the bump in $\sim 2 $ e-foldings,  the distance of the CMB scales to the end of inflation is only shifted by 2-efoldings with respect to the original model without the bump. This means that this modification of the original  potential will not alter its consistency with CMB data. 

We conclude that now the theory will produce BHs when the amplitude of $\zeta\sim 1/\beta \sim 0.17$, as opposed to the Gaussian case in which $\zeta_c \sim 0.54$. This difference corresponds to a factor 10 in the amplitude of the power spectrum\footnote{Note however that both scenarios are phenomenologically different and might lead to a  different spectrum of masses for the resulting PBH.}. Larger factors might be attained with a larger curvature at the local maxima, but the fine tuning in the potential keeps increasing.
    
\section{Conclusions}

In this paper we have calculated the amplitude and shape of non-Gaussianities in models of inflation where the field overshoots a local maximum. These models have gained much attention since they might create enough PBH to account for the dark matter in the Universe. The impact of non-Gaussianities on the abundance of PBH has so far only been estimated, in transient constant-roll backgrounds, through a perturbative determination of its amplitude and shape, more precisely, via the computation of the three-point function. Here we have shown that via the $\delta N$ formalism is possible to find a non-linear relation between a Gaussian random field $\zeta_g$ with the true curvature mode $\zeta$, given by eq. (\ref{eq:zetanptransf}). This relation incorporates the fact that large perturbations will be trapped in the false minimum, signaling the presence of a second critical amplitude $\zeta_*=1/\beta$. It also reduces to the standard local template of non-Gaussianities in the small perturbation regime.  We have computed the effect of such law in the case in which $\zeta_*$ is smaller or of the same order than the usual critical value for collapse of a large inhomogeneity. In the latter case, black holes will be created through the collapse of bubbles of false vacuum, rather than the usually considered collapse of a large overdensity. We have constructed a simple model in which this is the case, and for which amplitudes of the power spectrum 10 times smaller are needed in comparison with the Gaussian estimate in order to have a significant PBH abundance.

\acknowledgments{
We thank Jose J. Blanco-Pillado, Cristiano Germani, Ilia Musco and Sam Young for useful discussions. VA and JG are supported by FPA2016-76005 -C2-2-P, MDM-2014-0369 of ICCUB (Unidad de Excelencia Maria de Maeztu), AGAUR2014-SGR-1474 and AGAUR2017-SGR-754. AM-C acknowledges a postdoctoral contract from the Universidad de Cantabria postdoctoral program, the Spanish research project, ref. ESP2017-83921-C2-1-R (AEI/FEDER, EU), the Spanish Ministry MINECO grant (FPA2015-64041-C2-1P) and the Basque Government grant (IT-979-16).
}

\end{document}